\documentstyle[aps,multicol,pre]{revtex}
 \title{Thermodynamic non-additivity in disordered systems
}

 \author{E.V. Vakarin  and J.P. Badiali
}
 \address{Laboratoire de Electrochimie et Chimie Analytique,\\
  ENSCP-UPMC, 11 rue P. et M. Curie, 75231 Cedex 05, Paris, France\\
 }

\begin{document}
 \maketitle
 \begin{abstract}
   It is shown that there is a 
  mapping of the replica approach to disordered  systems with finite 
  replica index $n$ on the Tsallis non-extensive statistics, if the average 
  thermodynamic entropy differs from the information entropy for the 
  probability distribution. In the case of incomplete information
  the entropic index $q=1-n$ is shown to be related to the degree of lost 
  information. 
 \end{abstract}

\begin{multicols}{2}
 \section{Introduction}
 The non-extensive thermodynamics results from a 
 generalization\cite{Tsallis} of the standard Gibbs-Boltzmann (GB) 
 statistics.
 The non-extensive formalism 
  has been applied to the 
  description of a variety of physical systems\cite{site}. In some cases,
 related to disordered systems (e.g. anomalous
 diffusion in porous media or surface growth\cite{Spohn,Bukman}),
 it has resulted in quantitative implications, which are difficult
 to obtain in the framework of the classical GB approach.
 
 In the spirit of the {\it information theory} approach\cite{jaynes} 
 the non-extensivity is usually introduced starting with
 a postulation of the entropy measure. For instance, Tsallis 
 proposed\cite{Tsallis} the following form
 \begin{equation}
 S^T_q=\frac{1-\int (d\sigma) [p(\sigma)]^q}{q-1}
 \end{equation}
 where $p(\sigma)$ is a probability to find  a state $\sigma$,
 with $\int (d\sigma) p(\sigma)=1$. Here $q$ is an index, corresponding to 
 a given statistics. Simultaneously $q$ is a measure of the thermodynamic 
 non-additivity. Although in general the 
 non-extensivity should be distinguished\cite{touchette} from the 
 non-additivity, quite often these two terms are being used as synonymous.
 If the state $\sigma$ is characterized by the energy 
 $\varepsilon(\sigma)$, then, based on the entropy maximization under a 
 constraint for a generalized internal energy, one 
 obtains\cite{Curado,Mendes} a power law for the probability 
 \begin{equation}
 \label{p}
  p(\sigma)=\frac{[1-(1-q)\beta \varepsilon(\sigma)]^{1/(1-q)}}
  {\int (d\sigma) [1-(1-q)\beta \varepsilon(\sigma)]^{1/(1-q)}}
 \end{equation}
  We do not discuss here how the above distribution modifies with
  different constraints \cite{Mendes} for the internal energy as well as 
  the internal specificities \cite{Mendes} of the Tsallis 
 theory and its  mathematical consistency. 
  For any
  constraint\cite{Mendes} in the limit $q\to 1$ one recovers the standard GB 
  results: $S_{GB}=-\int(d\sigma)p(\sigma)\ln p(\sigma)$ and the exponential 
  form for $p(\sigma)$.  
  Note that the Tsallis form is not unique in having
  $S_{GB}$ as a limit. As an alternative example, one can consider the Renyi 
  form\cite{Renyi}, which is however additive. 
  
  Despite of many efforts\cite{Santos,Abe} towards deriving the Tsallis-type
  entropy starting from basic assumptions, up to now it is not clear which 
  physical systems are "naturally" non-extensive, and thus require the 
  Tsallis-type statistics as an essential tool. In particular,  
  the physical meaning of the $q$ parameter 
  \cite{Alemany,Beck,Wilk,Almeida} as well as a relation of $S_q^T$ to an 
  underlying  dynamics \cite{Cohen} 
  are still under discussion. Also the thermodynamic consistency of the 
  non-extensive formalism  has recently been criticized\cite{criticism}.  
   Therefore, a derivation of the power-law 
   entropy 
 (like $S_q^T$) instead of postulating it, would be much helpful in resolving
 these problems.   
 
 In this paper we investigate the conditions and assumptions 
 at which the non-extensive thermodynamics can be recovered using the tools 
 of the {\it standard statistical mechanics}, applied to systems which, 
 because of their complexity, require more than the usual Hamiltonian 
 description. This is in the spirit of the Beck-Cohen 
 approach\cite{BeckCohen}, who have shown that a non-GB distribution 
 (including the Tsallis one) could appear in systems with fluctuating 
 intensive quantities. Our purpose is to recover the Tsallis-type entropy, 
 which is a generating functional for the power-like probability 
 distribution, but not to construct the distribution itself. 
 This establishes a relation between the information
 theory and the statistical mechanics in application to systems which
 are out of the conventional GB equilibrium.
 In particular 
  we focus on disordered systems since their description involves two 
 essential ingredients: a Hamiltonian and a distribution function. 
 In fact, Tsallis has already commented\cite{Tsallisrep} on a formal 
 similarity of the non-extensive formalism and the replica trick, suitable 
 for the systems with a quenched disorder. Therefore it is tempting to 
 realize the condition at which such systems exhibit the 
 non-extensive behavior. This would allow us to make a link between the
 system dynamics (the Hamiltonian), its complexity (the probability 
 distribution) and the magnitude of the entropic $q$-parameter.

\section{Replica approach to disordered systems} 
From a dynamical point of view
disordered systems (like porous materials\cite{given} or spin 
glasses\cite{SK}) can be viewed as those consisting of many subsystems, 
whose equilibration times are much different\cite{Parisi1}. Then, 
considering the static (or 
 "equilibrium") properties, the slower subsystem, say $\{\sigma\}$, is 
 quenched with a given probability distribution $P(\sigma)$, while the 
  faster one, say $\{s\}$ is governed by a Hamiltonian 
 $H=H_R+ H[\sigma,s]$. In the case of spin glasses the
 counterparts of $\{s\}$ and  $\{\sigma\}$ systems are the spin variables
 and the random fields (or exchange constants), respectively. 
 It is essential that the Hamiltonian contains a reference part $H_R$, which
 gives a characteristic energy scale 
 reflecting the way of quenching (e.g. the cooling rate and the 
 correspondent temperature gap).   
 
  For a given 
  configuration $\{\sigma\}$ one can calculate the partition function
  \begin{equation}
  Z(\sigma)={\cal{Z}}/Z_R=\int (ds) e^{-\beta H[\sigma,s]}
  \end{equation}
  where $Z_R=e^{-\beta H_R}$ and $\beta=1/kT$. Then all 
  the thermodynamic characteristics are known. For instance, the 
  free energy excess
  \begin{equation}
  F(\sigma)={\cal{F}} - H_R=-\frac{1}{\beta} \ln Z(\sigma) 
  \end{equation}
  and the internal energy excess
  \begin{equation}
 \label{Usigma}
 U(\sigma)={\cal{U}}-H_R=-\frac{d}{d\beta}\ln Z(\sigma)
 \end{equation}
 determine the thermodynamic entropy
 \begin{equation} 
 S_T(\sigma)=\beta[U(\sigma)-F(\sigma)]
 \end{equation}
 Then all the relevant thermodynamic quantities can be obtained by averaging
 the quenched ones. For instance, the average free energy is given by
 \begin{equation}
 \label{free}
 F=\overline{F(\sigma)}=\int(d\sigma)P(\sigma)F(\sigma)  
 \end{equation}
 Note that, in contrast to the conventional equilibrium, now we have
 two entropies - the thermodynamic entropy
 \begin{equation}
 \label{STDN}
 S_T=\int(d\sigma)P(\sigma)S_T(\sigma)  
 \end{equation}
 and the one related to the information on the probability distribution
 \begin{equation}
 \label{SINF}
 S_I=-\int (d\sigma)P(\sigma)\ln P(\sigma)
 \end{equation}
  
 Instead of the direct averaging of the logarithm in (\ref{free}) it is 
 convenient to introduce the replica trick\cite{Mezard}, which is based on 
 the representation 
 \begin{equation}
 \label{replica} 
 \ln Z(\sigma)=\lim_{n\to 0} \frac{[Z(\sigma)]^n-1}{n}
 \end{equation} 
 It is equivalent to making $n$ noninteracting copies of the system.
 Nevertheless the copies are not completely independent because the
 probability distribution $P(\sigma)$ is 
 the same for all of them. Then the problem 
 reduces to the evaluation of the moments $\overline{[Z(\sigma)]^n}$, which 
 is expected to be a simpler task\cite{comment}. Note, however, that such a 
 procedure is not just a trick. In fact, the behavior of one system (e.g. a 
 spin glass) is interpreted as a limiting case of another system (the one 
 with $n \ne 0$). Such that the quenched average is expressed through a 
 "weighted" annealed average for a more complex system, involving
  an extension of the phase space\cite{Extended}.
 The systems with $n 
 \ne 0$ were considered \cite{Sherrington,Penney,Dotsenko} 
 in application to the spin glasses and neural networks\cite{neural}. In the 
 context of spin models such a system can be viewed as a collection of $n$ 
 domains of interacting spins with a common probability distribution 
 $P(\sigma)$. Note, however that due to the limiting procedure
  in eq.~(\ref{replica}) the replica index $n$ can be 
  non-integer\cite{comment}- i.e. at least for $0\le n \le 1$ it
  must vary continuously.   
 Sherrington\cite{Sherrington} has investigated the critical behavior
 of such replica magnets, driven by the inter- and intra-replica interaction.
 Moreover, Derrida 
 has shown \cite{Derrida}, that there is a mapping between the random walk, 
 the random energy model and the replica results for non-zero $n$.
  

Based on the discussion above we allow $n$ to be finite. 
The $\sigma$-dependent  
functions, like $F(\sigma)$, obey the standard statistical thermodynamics, 
since $\sigma$ is a parameter (although known only statistically at the 
level of $F(\sigma)$ evaluation). The average free energy is now 
$n$-dependent \begin{equation}
\label{Fn}
F_n=\overline{F_n(\sigma)}=-\frac{1}{\beta} 
\int (d\sigma) P(\sigma) \frac{[Z(\sigma)]^n-1}{n}
\end{equation} 
The internal energy $U_n$ can be found as a counterpart of the
 internal energy (\ref{Usigma})
and the average internal energy 
\begin{equation}
U_n=-\frac{1}{\beta}\int (d\sigma) P(\sigma)
[Z(\sigma)]^{n-1}dZ(\sigma)/d\beta
\end{equation} 
can be rearranged to the form
\begin{equation}
U_n=\int (d\sigma) P(\sigma) [Z(\sigma)]^n U(\sigma)
\end{equation}
If $\sigma$-dependent quantities obey the standard thermodynamic relation
$F(\sigma)=U(\sigma)-S_T(\sigma)/\beta$,
then the counterpart of the entropy is given by
\begin{equation}
\label{sn}
S_n=\int (d\sigma) P(\sigma)
\left[
\frac{[Z(\sigma)]^n(1+n \beta U(\sigma))-1}{n}
\right]
\end{equation}
 \section{Mapping on the Tsallis theory}
Starting from eq. (\ref{sn}) and replacing
 $n$ by $1-q$ we can rearrange the correspondent thermodynamic entropy $S_q$ 
 to the Tsallis construction 
 \begin{equation}
 \label{sq}
 S_q=\frac{1-\int (d\sigma) [\Pi(\sigma)]^q}{q-1}
 \end{equation}
 where $\Pi(\sigma)$ is given by
 \begin{equation}
 \label{PI}
 \Pi(\sigma)=
 \left[
  P(\sigma)[Z(\sigma)]^{1-q}\{1+(1-q)\beta U(\sigma) \} 
 \right]^{1/q}
 \end{equation}
 At this level
 our calculations are purely formal, and $S_q$ is equivalent to
 $S_n$. In order to obtain $S_q$ coherently with the Tsallis
 conjecture we have to associate $\Pi(\sigma)$ with a probability
 distribution. Then we naturally impose two restrictions:
 \begin{equation}
 \label{constraint}
 \Pi(\sigma)\ge 0; \quad \int (d \sigma)\Pi(\sigma)=1
 \end{equation}

  From the positivity condition we have
 \begin{equation}
 \label{constr1}
 q\le 1+ \frac{1}{\beta U(\sigma)}
 \end{equation}
 which is equivalent to the cut-off condition\cite{Mendes} in the
 Tsallis approach.
 
 In order to realize the meaning of $q \ne 1$ we expand $\Pi(\sigma)$,
 given by eq. (\ref{PI}), around $q=1$ up to the second order  
 \begin{eqnarray}
 \label{expansion}
 \Pi(\sigma)=
 P(\sigma)-P(\sigma)D(\sigma)(q-1)+\nonumber\\
 P(\sigma)
 \left[ D(\sigma)+
 \frac{1}{2} \left( D^2(\sigma)-\beta^2 U^2(\sigma) \right)
 \right](q-1)^2
 \end{eqnarray}
 where $D(\sigma)=S_T(\sigma)+\ln P(\sigma)$. Therefore, we see that
 the probability distribution differs from $P(\sigma)$ as long as
 $q \ne 1$. Moreover, the magnitude of $q$ is coupled to the 
 $\sigma$-dependent thermodynamics and to a shape of  $P(\sigma)$.
 
 From the 
 normalization condition we obtain
 \begin{equation}
 \label{eqforq}
 -\Delta S(q-1)+
 \left[\Delta S+\Sigma 
  \right](q-1)^2=0
 \end{equation}
 where
 \begin{equation}
 \Delta S=\int(d\sigma) P(\sigma) D(\sigma)=S_T-S_I
 \end{equation}
 is the difference of the thermodynamic and the information entropies and
 $\Sigma$ is given by
 \begin{equation}
 \Sigma=
 \frac{1}{2} \int(d\sigma) P(\sigma) 
 \left[ D^2(\sigma)-\beta^2 U^2(\sigma) \right]
 \end{equation}
 Then solving eq. (\ref{eqforq}) with respect to $q$ we obtain the result
 \begin{equation}
 \label{q}
 q-1=\frac{\Delta S}{\Delta S+\Sigma}
 \end{equation}
 suggesting that small deviations of $q$ from $1$ are determined by
 the difference $\Delta S$ of the thermodynamic and the information 
 entropies. Comparing eqs.(\ref{STDN}) and (\ref{SINF}) we can easily 
 demonstrate that these two entropic impacts are equal only if the 
 probability distribution is consistent with the thermodynamic 
 fluctuations\cite{fluct}: $P(\sigma) \propto \exp(-\beta S_T(\sigma))$. 
 Therefore, $\Delta S$ (and consequently $q$) can
 be viewed as a measure of the deviation from the conventional 
 thermodynamic equilibrium. Similar conjectures were drawn in 
 ref.~\cite{Vives}, discussing the fluctuations within the Tsallis 
 statistics. Since the sign of $\Delta S$ and $\Sigma$ is not fixed, then 
 from eq.~(\ref{q}) we can have both $q>1$ and $q<1$. Nevertheless, $q$ is 
 related to the replica index $n$ through $n=1-q$. Therefore, if $n \ge 0$
 (as it is usually assumed in the spin glass theories), then we deal
 with $q\le 1$.

 On the other hand $\Delta S$ depends explicitly on the thermodynamic
 parameters and on a shape of the quenching distribution.
 Therefore, there is a mapping of the replica approach
 with a non-zero replica index $n=1-q$ onto the Tsallis theory, provided
 that the replica index varies coherently with the system thermodynamics   
 and the probability distribution. This conclusion agrees with the results
 obtained within the incomplete information theory\cite{Wang} which suggests 
 that the $q$ parameter is related to the thermodynamic quantities. 
 In the case when the thermodynamic 
 parameters are fixed, the magnitude of $q$ is related to the parameters of
 the probability distribution. If, for instance, $P(\sigma)$ is 
 of the gaussian form, then the width of this distribution determines the 
 fluctuation $\overline {\sigma^2}-\overline {\sigma}^2 $. 
 In other 
 words the magnitude of $q$ is related to the fluctuation of the parameters
 in the quenching distribution. In some sense, this result is coherent
 with the one\cite{Wilk} obtained for the generalized exponential 
 distribution.   
 \section{Incomplete information}
As is discussed above, due to fact that $q \ne 1$, we deal with the 
weighted annealed averaging (see eq.~(\ref{sn})).  That is, in contrast to 
the usual quenching ($q=1$) we have a nonvanishing feed-back effect on 
$\{\sigma\}$-distribution from the thermodynamics of $\{s\}$-subsystem
(see eq.~(\ref{expansion})). 
As a result
 we operate with the probability distribution $\Pi(\sigma)$, involving the 
 thermodynamic quantities, instead of $P(\sigma)$ which corresponds to the 
 case when  $\{\sigma \}$-subsystem remains unchanged. This situation is 
 similar to what is reported\cite{Dotsenko,neural} for partially annealed 
 systems. 
  
  Taking the porous materials\cite{given} as an example, we can associate 
 $P(\sigma)$ with the empty matrix (e.g. pore size distribution) and 
 $\Pi(\sigma)$ - with the matrix in the presence of an adsorbed fluid. In 
 general, the latter changes the matrix properties. For instance, by the 
 analogy with the intercalation system\cite{PRB,JPCB}, we can expect a 
 change of the matrix volume upon the fluid absorption. Moreover, quite 
 often the adsorbates induce a new ordering\cite{PRL} of the substrates or 
 facilitate their disordering\cite{PRBr} depending upon the specificities of 
 the adsorbate-substrate interaction.
 Therefore, because of a 
 complicated fluid-matrix coupling, our information on the pore size 
 distribution could be incomplete\cite{Wang} 
 \begin{equation}
 \label{incomp}
 \int (d\sigma) P(\sigma)=1; \quad \int (d\sigma) \Pi(\sigma)=\Lambda \ne 1
 \end{equation}
 where $1-\Lambda$ can be considered as a degree of lost information. In 
 this case we can rearrange (\ref{sn}) to the following form
 \begin{equation}
 \label{sqn}
 S_q=-\int (d\sigma) \Pi(\sigma) \frac{[\Pi(\sigma)]^{q-1}-1}{q-1}
 \end{equation}
 where $\Pi(\sigma)$ should be found from (if $q \ne 1$)
 \begin{eqnarray}
 [\Pi(\sigma)]^{q}-\Pi(\sigma)=\nonumber\\
 P(\sigma)\left[[Z(\sigma)]^{1-q}\{1+(1-q)\beta U(\sigma) \} 
 -1 \right]
 \end{eqnarray}
 For small $|1-\Lambda|$ the distribution $\Pi(\sigma)$ should exhibit
 small deviations from that given by eq.~(\ref{PI}). Therefore, we determine
 $\Pi(\sigma)$ perturbatively, taking eq.~(\ref{PI}) as a zeroth 
 approximation. Then, expanding around $q=1$, we find that the linear  
 approximation is the same as that in eq.~(\ref{expansion})
 \begin{equation}
 \label{expansion1}
 \Pi(\sigma)=
 P(\sigma)-P(\sigma)D(\sigma)(q-1)
 \end{equation} 
 From the incomplete normalization condition (\ref{incomp})
  we obtain   
 \begin{equation}
 q=1+\frac{1-\Lambda}{\Delta S}
 \end{equation}
 which demonstrates that the entropic index $q$ is related to the 
 degree of lost information ($1-\Lambda$). In the limit $\Lambda \to 1$
 we must retain the quadratic term in (\ref{expansion1}) and then we recover
 our previous result (\ref{q}). 
 \section{Conclusion}
 A relation between the non-extensive Tsallis statistics and the replica
 formalism for disordered systems is discussed.
 The key specificity
 of this kind of systems is the splitting into subsystems with different
 levels of description, i.e. the Hamiltonian and the probability 
 distribution. Due to this one necessarily operates with two entropic impacts:
 the thermodynamic and the information entropies. 
 It is shown that there is a mapping of the 
 replica approach to disordered systems on the Tsallis theory with $q \ne 1$, 
 if the average thermodynamic entropy differs from the information
 entropy related to the probability distribution. In the case of incomplete
 information the entropic index deviates from unity  proportionally to the 
 degree of lost information. 
 Therefore, the 
 magnitude of $q$ depends on the thermodynamic parameters as well as on a 
 shape of the distribution. This 
 conclusion is coherent with the results of other studies 
 \cite{Wilk,BeckCohen,Vives,Wang} discussing the meaning of the $q$ index in 
 the Tsallis approach.  
 
 Our results suggest that a similar mapping can be expected for
 other systems which are out of the thermodynamic equilibrium (at least in 
 its traditional meaning). Then the problem involves additional 
 relevant quantities, related to a deviation from the conventional
 equilibrium. For instance, the deviation could be induced by the heat bath 
 properties \cite{Almeida}, by fluctuations of intensive thermodynamic 
 parameters\cite{BeckCohen}, or by finite sizes (no thermodynamic 
 limit)\cite{Adib}. 
 
 It should be emphasized that the non-additive nature of the system 
 discussed here does not depend on the approximations made. From the every 
 formulation of the problem we have two entropies $S_I$ and $S_T$. If we 
 consider a "total" entropy $\Omega $ as a measure of our uncertainty
 on the system state, then
 it is clear that $\Omega$ cannot be written as an additive combination
 of $S_I$ and $S_T$ because these two are not independent. Therefore it 
 would be interesting to find a way of solving this more general problem
 without introducung model arguments.

\section{Acknowledgment} 
The authors are much grateful to A.~Le~Mehaute, L.~Nivanen and Q.~A.~Wang
for helpful discussions.     

\end{multicols}

\begin{thebibliography}{99}
\bibitem{comment}
 Here we do not discuss the problems of the replica theory itself, such 
 as the limiting procedure or the replica symmetry breaking.
\end{thebibliography}
\end{document}